\documentclass[usenatbib]{mnras}
\usepackage{aasmacros}
\usepackage{amsmath,times}
\usepackage{graphicx,url}
\usepackage{amssymb}
\usepackage[usenames,dvipsnames]{color}
\usepackage{multirow}
\usepackage{natbib}
\usepackage{siunitx}
\usepackage{booktabs}
\usepackage{placeins}
\usepackage{xspace}
\usepackage[normalem]{ulem}
\usepackage{float}
\usepackage{appendix}

\newcommand{\msun}{$\mathrm{M}_\odot$\xspace}

\newcommand{\auriga}{\textsc{Auriga }}
\newcommand{\apostle}{\textsc{Apostle }}
\newcommand{\aquarius}{\textsc{Aquarius }}
\newcommand{\rvir}{$r_{200}$\xspace}
\newcommand{\mvir}{$M_{200}$\xspace}

\title{Subhalo destruction in the \apostle and \auriga simulations}

\author[Richings et al.]{Jack Richings$^{1,2}$\thanks{Contact e-mail: 
\href{jack.richings@durham.ac.uk}{jack.richings@durham.ac.uk}}, Carlos Frenk$^1$, Adrian Jenkins$^1$, Andrew Robertson$^1$, 
\newauthor 
Azadeh Fattahi$^1$, Robert J. J. Grand$^3$, Julio Navarro$^4$, R{\"u}diger Pakmor$^{3,5}$
\newauthor
Facundo A. Gomez$^{3,7}$, Federico Marinacci$^{8,9}$
\\
$^1$ Institute for Computational Cosmology, Department of Physics, University of Durham, South Road, Durham DH1 3LE, UK\\
$^2$ Institute for Particle Physics Phenomenology, Department of Physics, University of Durham, South Road, Durham DH1 3LE, UK\\
$^3$ Max-Planck-Institut fur Astrophysik, Karl-Schwarzschild-Str. 1, D-85748 Garching, Germany\\
$^4$ Department of Physics and Astronomy, University of Victoria, PO Box 3055 STN CSC, Victoria, BC, V8W 3P6, Canada\\
$^5$ Heidelberger Institut fur Theoretische Studien, Schloss-Wolfsbrunnenweg 35, 69118 Heidelberg, Germany\\
$^6$ Instituto de Investigacion Multidisciplinar en Ciencia y Tecnologia, Universidad de La Serena, Raul Bitran 1305, La Serena, Chile\\
$^7$  Departamento de Fisica y Astronomia, Universidad de La Serena, Av. Juan Cisternas 1200 N, La Serena, Chile\\
$^8$ Kavli Institute for Astrophysics and Space Research, Massachusetts Institute of Technology, Cambridge, MA 02139, USA\\
$^9$ Harvard-Smithsonian Center for Astrophysics, 60 Garden Street, Cambridge, MA 02138, USA
}

\begin{document}

    \pubyear{2019}

  \maketitle

  \label{firstpage}

\begin{abstract}
  N-body simulations make unambiguous predictions for the abundance of
  substructures within dark matter halos. However, the inclusion of baryons
  in the simulations changes the picture
  because processes associated with the presence of a large galaxy in
  the halo can destroy subhalos and substantially alter the mass
  function and velocity distribution of subhalos. We compare the
  effect of galaxy formation on subhalo populations in two state-of-the-art sets of
  hydrodynamical $\Lambda$CDM simulations of Milky Way mass halos,
  \apostle and \textsc{Auriga}. We introduce a new method for tracking
  the orbits of subhalos between simulation snapshots that gives
  accurate results down to a few kiloparsecs from the centre of the
  halo. Relative to a dark matter-only simulation, the abundance of
  subhalos in \apostle is reduced by 50\% near the centre and by 10\%
  within \rvir. In \auriga the corresponding numbers are 80\% and
  40\%. The velocity distributions of subhalos are also affected by
  the presence of the galaxy, much more so in \auriga than in
  \apostle. The differences on subhalo properties in the two
  simulations can be traced back to the mass of the central galaxies,
  which in \auriga are typically twice as massive as those in
  \apostle. We show that some of the results from previous studies are
  inaccurate due to systematic errors in the modelling of subhalo orbits near
  the centre of halos.
\end{abstract}

\begin{keywords}
cosmology: theory -- cosmology: dark matter -- methods: N-body simulations -- galaxies: kinematics and dynamics

\end{keywords}

\section{Introduction}
\label{Intro}
In the $\Lambda$-Cold Dark Matter ($\Lambda$CDM) model of cosmology,
the formation of cosmic structure proceeds hierarchically by the
merging of smaller structures to form larger ones \citep{peebles1980,
  davis1985}. Whilst the merging process is incomplete, substructures
can survive within the dark matter halo of a galaxy or cluster
\citep{ghigna1998}. In galaxies like the Milky Way many more such
substructures survive than there are visible satellites
\citep{moore1999,klypin1999}. This disparity is the natural outcome of
physical processes known to be important in galaxy formation: the
reionization of hydrogen around redshift $z=8$ \citep{planck2016} and
the expulsion of gas heated by supernovae \citep{bullock2000,
  benson2002, somerville2002, okamoto2008, sawala2016, maccio2010}.
Similarly, an apparent absence of visible galaxies in the most massive
subhalos that form in $\Lambda$CDM dark-matter-only simulations
\citep{boylan2011} can be readily explained by processes related to
gas expulsion from subhalos at early times
\citep{sawala2013,sawala2016}.

Even though baryon effects are sufficient to account for the abundance
of galactic satellites within the standard $\Lambda$CDM model, a
number of alternative models for the nature of the dark matter have
been proposed motivated largely by a desire to explain these so-called
``missing satellites'' and ``too-big-to-fail'' problems. \citep[e.g][]
{spergel2000,colin2000,petraki2013,jascha2015,hui2017}. With a
judicious choice of the additional parameters in these alternative
models, e.g. the mass of a warm dark matter (WDM) particle, the
abundance of satellites in the Milky Way can also be reproduced
\citep{lovell2012,lovell2017}. A particularly interesting WDM
candidate is motivated by the discovery of a 3.5~keV emission line in
the X-ray spectra of galaxies and clusters \citep{bulbul2014,
  boyarsky2014}. Whilst the nature of the origin of this line is
disputed \citep{malyshev2014, anderson2015, jeltema2015,franse2016,
  sorensen2016}, if its origin is not explicable within the standard
  model of particle physics, it could be the
result of the decay of 7 keV sterile neutrino dark matter.

A key prediction that distinguishes CDM from some of the alternatives,
such as WDM, is the abundance of small-mass halos and subhalos.  In
CDM, the halo mass function continues to rise to small masses
\citep{diemand2007,springel2008}, whereas in WDM, the halo mass
function is truncated at a mass on the scale corresponding to
dwarf galaxies \citep{colin2000, lovell2012, schneider2012, hellwing2016,
  bose2017}. In sterile neutrino models, the power spectrum of
primordial fluctuations depends not only on the dark matter particle
mass but also on an additional lepton asymmetry parameter. In the
coldest sterile neutrino model compatible with the 3.5~keV line
originating from particle decay, the mass function is suppressed by a
factor of 5 relative to CDM at $10^8$ \msun and is negligible at
$10^7$ \msun\citep{bose2017}. Thus, detection of halos of mass below
$10^7$ \msun would rule out this candidate particle and set a lower
limit larger than 7 keV for the sterile neutrino mass. Conversely, a
convincing non-detection of halos of mass below $\sim10^8$ \msun would
rule out CDM \citep{li2016}.

If they exist, the vast majority of these small-mass halos will be
dark, that is, almost completely devoid of baryonic matter. This
baryon deficit is the result of reionisation and supernova heating
\citep{okamoto2008,sawala2016}. These dark objects can be detected
through their gravitational interaction with visible matter. A
particularly promising test is gravitational imaging
\citep{koopmans2005} in which low-mass halos perturb the giant arcs or
Einstein rings that can be produced when a background galaxy is
strongly lensed. This method has already yielded detection of a
$1.9 \pm 0.1$ $\times 10^8$ \msun dark satellite and, with imaging
data of good quality, the detection sensitivity could reach
2$\times 10^7$ \msun \citep{vegetti2012}\footnote{The definition of
  mass here is based on a pseudo-Jaffe model and differs from the
  standard definition of halo and subhalo masses used in cosmological
  simulations and in this
  paper.}\\

Although for practical lensing configurations the lensing signal is
dominated by field halos rather than subhalos \citep{li2017,
  despali2018}, the latter make a non-negligible contribution to the
lensing distortion.  Since dark subhalos in this low-mass range are
uncontaminated by baryonic matter at the present day, the only
uncertainty in their abundance arises from possible interactions
between subhalos and the central galaxy in their common host halo, for
example tidal disruption. Quantifying these effects is necessary to 
make accurate predictions for the expected lensing signals.

The abundance of dark substructure in our own Galaxy may be probed in
other ways. For example, stellar streams, formed by the tidal
disruption of globular clusters or dwarf galaxies, can be measurably
perturbed by passing substructures which produce gaps in the streams
\citep{carlberg2012}. Surveys such as \textsc{Gaia}
\citep{perryman2001, gilmore2012}, \textsc{DES} \citep{des2005}, and
\textsc{LSST} \citep{lsst2009} have the potential to measure these
gaps and thereby determine the mass function of substructures in the
Milky Way down to a scale of $10^7$\msun \citep{erkal2015a,
  erkal2015b}. Such methods were explored in \citet{erkal2016}; their 
results are affected by a number of uncertainties, as the simulations
used did not incorporate baryonic physics, and a particular velocity
distribution of subhalos was assumed to break the degeneracy in the
method between peturber mass and velocity.

The role of the central galaxy in the destruction of substructure has
been studied using N-body simulations that incorporate an analytic
disk potential \citep{donghia2010,yurin2015}, as well as
hydrodynamical simulations \citep{garrison2017,sawala2017}. The specific
implementation of baryonic physics is important: the choice of subgrid
model, physical parameters and method for solving the hydrodynamical 
equations all individually can affect the abundance of substructure. \citet{errani2017} 
also showed that the inner slope of the density profile of infalling substructures
affects their survival probability. \citet{benitez2018} showed that the central
density of dwarf galaxies depends strongly on the choice of the star formation gas 
density threshold, a CDM cosmological simulation producing cuspy or cored profiles
depending on the choice of this parameter.

The effect of changing the subgrid galaxy formation models on subhalo
abundance has been investigated by \citet{despali2017} in the case of
the \textsc{Eagle} and \textsc{\textsc{Illustris}} 100$^3$ Mpc$^3$
simulations \citep{schaye2015, vogelsberger2014}. Both simulations
have relatively poor mass resolution (approximately $10^7$ \msun) so
this study was restricted to massive substructures rather than the
small ones that are important for distinguishing CDM from
WDM. Furthermore, these simulations have a relatively small number of
outputs so the orbits of subhalos cannot be tracked and, as a result,
the destruction of subhalos in the innermost regions of galaxies,
where processes such as disk shocking are important, is poorly
sampled.

With mass resolution of approximately $10^4$ \msun, the simulations
that we analyze in this paper have at least 100 times better
resolution than the simulations studied by \citet{despali2017}. In
particular, they resolve the small-mass halos (mass $\sim 10^7$ \msun)
required to distinguish CDM from WDM. To investigate the dependence of
the surviving subhalo abundance on the choice of baryonic physics
implementation, we compare the \apostle \citep{sawala2016, fattahi2016} and \auriga
\citep{grand2016} CDM simulations. We integrate the orbits of subhalos
between snapshots to obtain precise estimates of time-averaged subhalo
abundance close to the centre of the halo. This is the first direct
comparison of baryonic physics models at such a high level of
resolution, both spatially and temporally.

\section{Methods}
\label{Methods}

\subsection{Simulations}
\label{Simulations}

We use two suites of simulations to study the impact of baryons on
galactic substructure. The first is a set of zoom simulations of
Local Group-like volumes from the \apostle project \citep{fattahi2016,
  sawala2016}. Each volume contains a pair of halos, each of mass
$\sim10^{12}$ \msun, corresponding to the Milky Way and Andromeda.  We
study the same two volumes considered by \citet{sawala2017}, giving a
total of four high-resolution halos. The second suite of simulations,
taken from the \auriga project \citep{grand2016}, is a set of zoom
simulations of individual Milky-Way sized galaxies, selected from the
\textsc{Eagle} 100$^3$ Mpc$^3$ simulation (L0100N1504)
\citep{schaye2015}. There are six high-resolution (``level~3'') halos
in the \auriga sample. In addition, in \S~\ref{LargeRadii} we
also analyze the larger , ``level~4'', sample of 30 halos simulated at 10
times lower mass resolution than the level~3 examples.

For each simulation we have both a dark matter only (DMO) version and
a version including baryonic physics relevant to galaxy formation (gas
cooling, star formation, chemical enrichment, black hole formation,
feedback from stellar evolution, AGN, etc.)  The \apostle simulations
were performed with with the \textsc{Eagle} reference model
\citep{schaye2015, crain2015}, which is based on Gadget3,
while the \auriga simulations were performed with a variant of the
Arepo code \citep{springel2011,grand2016} used for the
\textsc{Illustris} simulation \citep{vogelsberger2014}. The parameters
of the subgrid models in \textsc{Eagle} and \textsc{Illustris} are
calibrated somewhat differently. In \textsc{Eagle}, they are chosen so
as to reproduce the $z=0$ galaxy stellar mass function and size
distribution, while in \textsc{Illustris} they are tuned to match the
$z=0$ ratio of galaxy stellar to dark matter mass and the cosmic star
formation rate at all times. Key diagnostics of each halo, as well as
relevant simulation parameters are listed in Table~\ref{HaloTable}.\\ 

The main halos in the \apostle and \auriga simulations have broadly
similar masses, $\sim 10^{12}$~\msun; however, the stellar masses of
the central galaxies in \auriga are significantly larger, typically
around twice as massive as an \apostle galaxy. The \auriga galaxies
are also more concentrated than the \apostle galaxies; despite being
twice as massive, their half-stellar-mass radii are similar or smaller
than those of \apostle galaxies.

\begin{table*}
\resizebox{\textwidth}{!}{%
\begin{tabular}{@{}lccccccc@{}}
\toprule
\multicolumn{1}{c}{}            & \multicolumn{2}{c}{M$_{200}$ {[}$10^{12}$\msun{]}} & \multicolumn{2}{c}{N$_{\mathrm{sub}}$}         & M$_{\mathrm{gal}}$ {[}$10^{10}$\msun{]} & $m_{\mathrm{DM}}$ {[}$10^{4}$\msun{]} & Softening {[}kpc{]} \\ \cmidrule(l){2-8} 
\multicolumn{1}{c}{}            & \multicolumn{1}{c|}{DMO}     & Hydro     & \multicolumn{1}{c|}{DMO} & Hydro &                      &                    &                \\ \midrule
\multicolumn{1}{l|}{\apostle- 11} & 1.66                         & 1.57      & 2027                     & 1543  & 3.40                 & 4.92               & 0.13           \\
\multicolumn{1}{l|}{\apostle- 12} & 1.10                         & 1.02      & 2158                     & 1563  & 1.87                 & 4.92               & 0.13           \\
\multicolumn{1}{l|}{\apostle- 41} & 1.35                         & 1.16      & 1579                     & 1253  & 1.11                 & 2.45               & 0.13           \\
\multicolumn{1}{l|}{\apostle- 42} & 1.39                         & 1.12      & 2650                     & 1675  & 1.94                 & 2.45               & 0.13           \\
\multicolumn{1}{l|}{\auriga- 6}   & 1.05                         & 1.01      & 1355                     & 675  & 7.30                 & 3.64               & 0.18           \\
\multicolumn{1}{l|}{\auriga- 16}  & 1.49                         & 1.50      & 2088                     & 936  & 9.28                 & 3.89               & 0.18           \\
\multicolumn{1}{l|}{\auriga- 21}  & 1.51                         & 1.42      & 2094                     & 932  & 11.23                 & 4.11               & 0.18           \\
\multicolumn{1}{l|}{\auriga- 23}  & 1.61                         & 1.50      & 1870                     & 915  & 10.45                 & 4.35               & 0.18           \\
\multicolumn{1}{l|}{\auriga- 24}  & 1.50                         & 1.47      & 1982                     & 943  & 9.12                 & 3.92               & 0.18           \\
\multicolumn{1}{l|}{\auriga- 27}  & 1.76                         & 1.70      & 2351                     & 1021  & 10.99                & 4.03               & 0.18           \\ \bottomrule
\end{tabular}%
}
\caption{Properties of the halos analyzed in this work at redshift
  $z=0$. Each \auriga halo is the largest object identified using the friends-friends-algorithm. Each \apostle halo is either the largest or second largest friends-of-friends group. N$_{\mathrm{sub}}$ is the number of subhalos identified by
  the \textsc{Subfind} algorithm \citep{springel2001}, with mass
  greater than 10$^{6.5}$   \msun, inside \rvir; M$_{\mathrm{gal}}$ is 
  the total mass of all gas
  and star particles within 30 kpc from the centre of the
  halo; $m_{\mathrm{DM}}$ is the mass of the high-resolution dark
  matter particle used in the hydrodynamical simulations. The
  softening is the value appropriate to the high-resolution dark 
  matter particles at redshift $z=0$.} 
\label{HaloTable}
\end{table*}

All quantities in this paper are averaged over the 5 Gyr period,
between redshift $z=0.5$ and the present day, to give an expected
probability density over this time interval. In both simulations, halos
are identified using the friends-of-friends algorithm \citep{davis1985}.
Halo substructure is identified using the \textsc{Subfind} algorithm \citep{springel2001}.
When computing averages over multiple halos, we take the median value in physical units. In
Table~\ref{HaloTable} and throughout this work we define the size and
mass of a host halo through its virial radius, \rvir, and mass, \mvir,
respectively. The virial radius, \rvir, is defined as the radius
within which the mean mass density is 200 times the critical density
of the universe. The virial mass, \mvir, is the total mass
enclosed within that radius. When performing time-averaged calculations which
span several snapshots, we interpolate \rvir and \mvir linearly in time.

\subsection{Halo masses}
\label{MassCorrection}
In DMO simulations, the baryonic mass is collisionless: around
15\% (the value of $\Omega_b/\Omega_m$) of the mass of each simulation
particle represents ``collisionless baryons''. In hydrodynamical
simulations, low-mass halos lose much of their baryonic mass during
reionisation or, subsequently, through galactic winds powered by
supernovae. DMO halos cannot undergo this mass loss, and so they will
be approximately 15\% more massive than their hydrodynamical 
counterparts at early times. This difference in mass is exacerbated
with time because more massive halos accrete mass at a higher rate
than smaller mass halos and thus grow faster. For an isolated
$10^8$~\msun halo at redshift $z=0$, this difference in mass between
the same object with hydrodynamics or DMO is typically around 20-30\%
\citep{sawala2013,sawala2016}.

Most of the results presented in this work do not include a correction
for this effect. This is largely in order to make it easier to compare
with previous studies based on DMO simulations. However,
if we wish to identify what fraction of the reduction in halo
abundance is attributable to interactions with the host galaxy rather
than to  this environmentally-independent mass loss effect, it is
necessary to correct the mass of DMO halos. We use this correction
for some of the calculations presented in \S~\ref{LargeRadii}.\\

\begin{figure}
  \includegraphics[width=\linewidth]{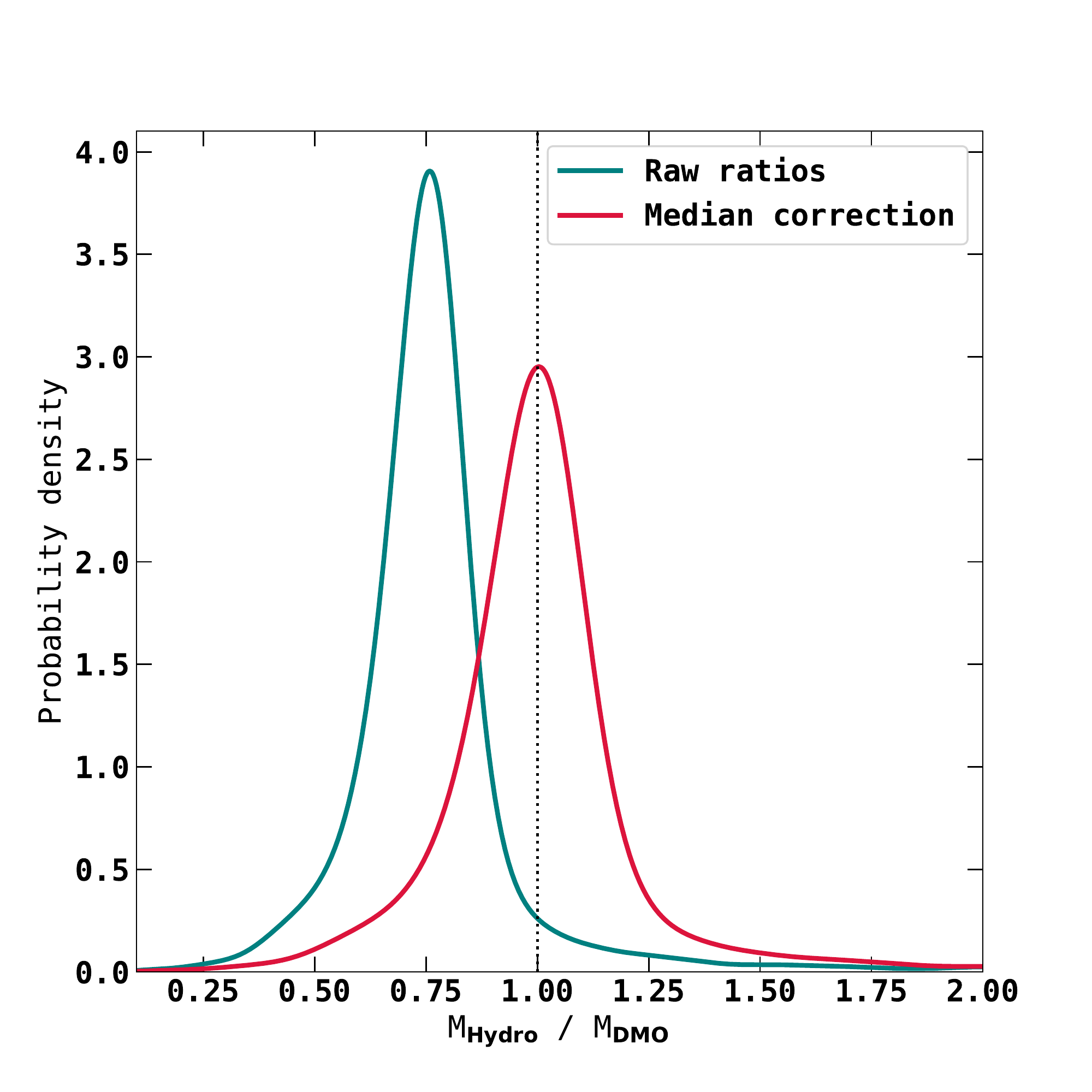}
  \caption{Distribution of the ratios of the masses of halos matched
    between DMO and hydrodynamical versions of a simulation. The teal
    line shows the distribution of mass ratios when no correction has
    been applied. The crimson line shows the distribution of mass
    ratios after the masses of DMO halos have been multiplied by
    the median of the teal distribution (a value of 0.76).}
  \label{fig:mfac_dist}
\end{figure}

The procedure we use to correct DMO halo masses is as follows. We
match halos between the DMO and hydrodynamical versions of a
simulation using the particle matching criteria of \citet{bose2017},
in which the 50 most bound DM particles of halos are matched
bijectively between the DMO and hydrodynamical simulations. We then
form a matched ``field'' sample by selecting halos which are at least
500 kpc from a galaxy in the hydrodynamical version of the simulation,
so as to avoid any differences due to evolution in the tidal field of
the main halo. For each \auriga level 3 volume, we have approximately 1000
matched objects with mass between $10^7$--$10^8$ \msun. The numbers for \apostle
are significantly larger as a greater fraction of the simulation is field volume.
For each pair of matched halos we calculate the ratio of their masses. We take
a DMO halo's ``effective mass'' to be the mass assigned to it by the
\textsc{Subfind} algorithm, multiplied by the median of the distribution of mass
ratios of this matched sample. The distributions of mass ratios before
and after this procedure are shown in Fig.~\ref{fig:mfac_dist}.  When the masses
of DMO subhalos are corrected by the median mass ratio, the peak of the mass
ratio distribution will occur at a value of 1, by construction (this would not
be the case if we had corrected by the mean mass ratio). The width of the corrected
distribution is around 30\% larger for the corrected distribution. The results
shown in Fig. \ref{fig:mfac_dist} are calculated using
only subhalos at redshift $z=0$. We have checked that for
redshifts between $z=1$ and the present day, the size of this effect 
is independent of redshift.

We find that the correction factor has no dependence on mass for halos
with DMO masses between $10^7-10^9$ \msun. For the \auriga simulations, we 
find a median correction factor of 0.76, and the interquartile range of
of correction factors is 0.12. For the \apostle simulations the median
correction factor is 0.75, due to the slightly different choice of cosmological
parameters in the simulation. We find that this
correction procedure does not work well for halos with masses below
$10^7$ \msun. The probability of a halo being matched between
simulations falls steeply for halos smaller than this.  Furthermore,
the distribution of mass ratios will be biased as the resolution limit
of the simulation imposes a limit on the smallest possible mass ratio.
Therefore, when correcting halo masses, we restrict our attention to
halos with masses greater than $10^7$ \msun.

\subsection{Orbits}

The time between snapshots in the simulations (around 300 Myr for
\apostle and less for \auriga) is greater than the crossing time for
the central 20-30 kpc of the main halo. These snapshots are sufficiently
infrequent that the subhalo abundance in the central 20 kpc of the halo is
poorly sampled. To make precise theoretical predictions for the
abundance of substructure near the centre of halos and to quantify the
impact of the galactic disk, previous studies
inferred the positions of subhalos between snapshots using a cubic
spline to interpolate between snapshots \citep{sawala2017,
  garrison2017}. Specifically, a cubic piecewise polynomial was fit to
each Cartesian coordinate of the physical positions of subhalos at the
snapshots as a function of time with the condition that the result be
twice continuously differentiable, except at the ends, where the first
derivative is equal to the linear interpolant slope.

We show in \S\ref{spline} that this method is biased. Cubic spline interpolation
systematically underpredicts the orbital radii of subhalos at
distances of less than 30 kpc from the centre of the halo, precisely
the region where reconstructing subhalo orbits is most important for
tests of the CDM model. Orbital radii are often underpredicted by a
factor of two or more, especially if pericentre occurs at at time
halfway between two snapshots.

Instead of interpolating, we track the positions and velocities of
subhalos between snapshots by integrating their orbits in the
potential of the halo, which we assume to be static over this time
and, for simplicity, axisymmetric. We model the potential and
integrate the orbits using the publicly available codes \textsc{Galpy}
and \textsc{Pynbody} \citep{bovy2015,pynbody}. This method accurately
reproduces the orbits of subhalos around the host halo, even in
situations where the cubic spline method is most prone to failure. By
integrating the orbits of subhalos we can accurately estimate subhalo
abundances at galactic distances of less than 10 kpc.

To predict the position of a subhalo accurately, choosing the
correct frame of reference is paramount. Following the prescription of
\citet{Lowing2011} we take the coordinate origin of the halo to be the
position of the particle with the minimum potential energy, and the velocity of the parent
halo (which is to be subtracted from the velocity of the subhalo under
consideration) to be the mean velocity of all particles within 5\% of
\rvir. We define this reference frame for each snapshot. All
calculations are performed in physical coordinates.

\begin{figure}
  \includegraphics[width=\linewidth]{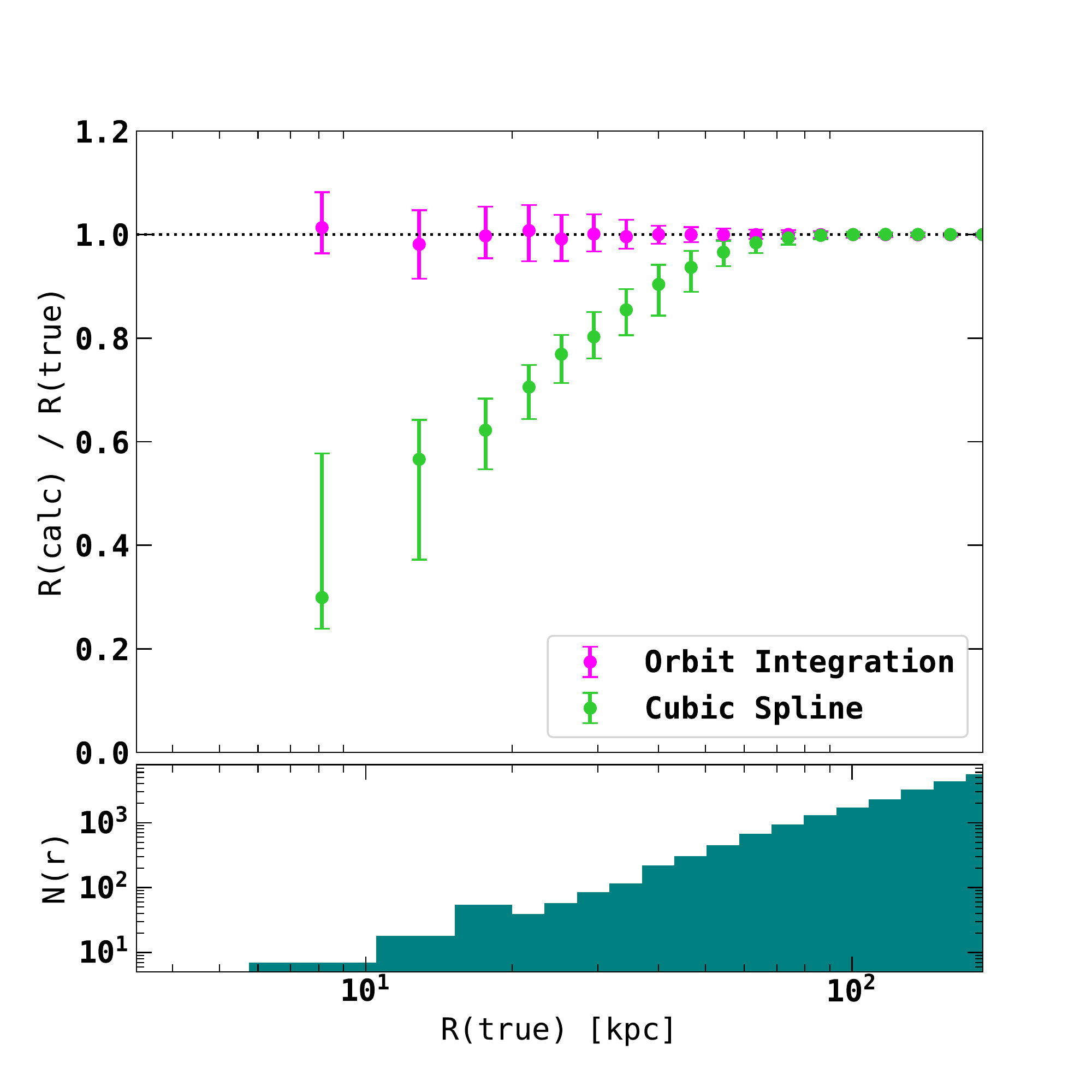}
  \caption{The accuracy of orbit integration at predicting the orbital
    radii of subhalos at an intermediate snapshot. For reference, we
    compare this method to the cubic spline described in
    \citet{sawala2017} and also used in \citet{garrison2017}. The
    lower panel shows the number of subhalos present in each radial
    bin.}
  \label{fig:leapfrog_test}
\end{figure}

We match subhalos between snapshots using a merger tree. To determine
the position and velocity of a subhalo between snapshots 1 and 2, in
the time interval $t_1<t<t_2$, we take the following steps: 
\begin{enumerate}

 \item Construct an intermediate ``snapshot'' by summing the mass
   distributions of snapshots 1 and 2, halving the mass of each
   particle. 

 \item Since the required \textsc{Galpy} routines are written for
   axisymmetric potentials, we interpolate the mass distribution of
   the intermediate snapshot on a 2-dimensional $R-z$
   grid\footnote{$R$ is the two-dimensional cylindrical radius, and $z$
     is the vertical distance.}. We discard particles which are further
     than 800 kpc from the centre of the halo. The effect of this approximation
     on the calculated orbits is negligible. The $z$-axis of the grid is
   taken to be the $z$-direction in simulation coordinates, and so is
   unrelated to the plane of the galaxy. The accuracy of the results
   in Fig.~\ref{fig:leapfrog_test} shows that this arbitrary choice of
   $z$-axis is unimportant as the mass distribution is close to
   spherical. 

 \item Taking the subhalo at snapshot 1 to be a point mass, integrate
   its orbit forwards in time in the intermediate potential using the standard \textsc{Galpy}
   fourth-order symplectic integrator. 

 \item Integrate the orbit of the subhalo at snapshot 2 backwards in
   time in the intermediate potential.

 \item The orbit of the  subhalo is found by taking a weighted sum of
   the forwards and backwards orbits. The position, $\vec{x}$, of a
   subhalo at a time $t$ in the interval $t_1<t<t_2$ is given by:
 \begin{equation}
     \vec{x}(t) = \vec{x}_f(t)\frac{t_2-t}{t_2-t_1} +
     \vec{x}_b(t)\frac{t-t_1}{t_2-t_1}, 
 \end{equation}
 where $\vec{x}_f$ and $\vec{x}_b$ are the positions of the subhalos
 being integrated forward and backward in time at time $t$
 respectively. 
 \item The position and velocity of each subhalo is output every 3~Myr.
\end{enumerate}

To assess the accuracy of the reconstruction of subhalo orbits we
perform the following experiment. We select a pair of non-consecutive
snapshots from the \auriga simulation (snapshots 99 and 101,
corresponding to a redshift of $z\simeq 0.4$). The time between these
snapshots is approximately the same as the time between successive
\apostle snapshots. Using our method, we calculate the positions of
all subhalos at the time of snapshot 100, which we can compare
directly with the actual positions calculated at the intermediate
snapshot. The results of this test are shown in
Fig.~\ref{fig:leapfrog_test}. We can see that orbit integration
accurately predicts the positions of subhalos between snapshots, and
is therefore an effective tool for studying the dynamics of
substructure close to the centre of the halo. The green points in
Fig. \ref{fig:leapfrog_test} show the results of the same test when
applied to subhalo orbits calculated using the cubic spline method.
A detailed study of why the cubic spline method underpredicts the orbital
radii of subhalos is given in the following subsection.

\subsubsection{Comparison to cubic spline}
\label{spline}

\begin{figure}
  \includegraphics[width=\linewidth]{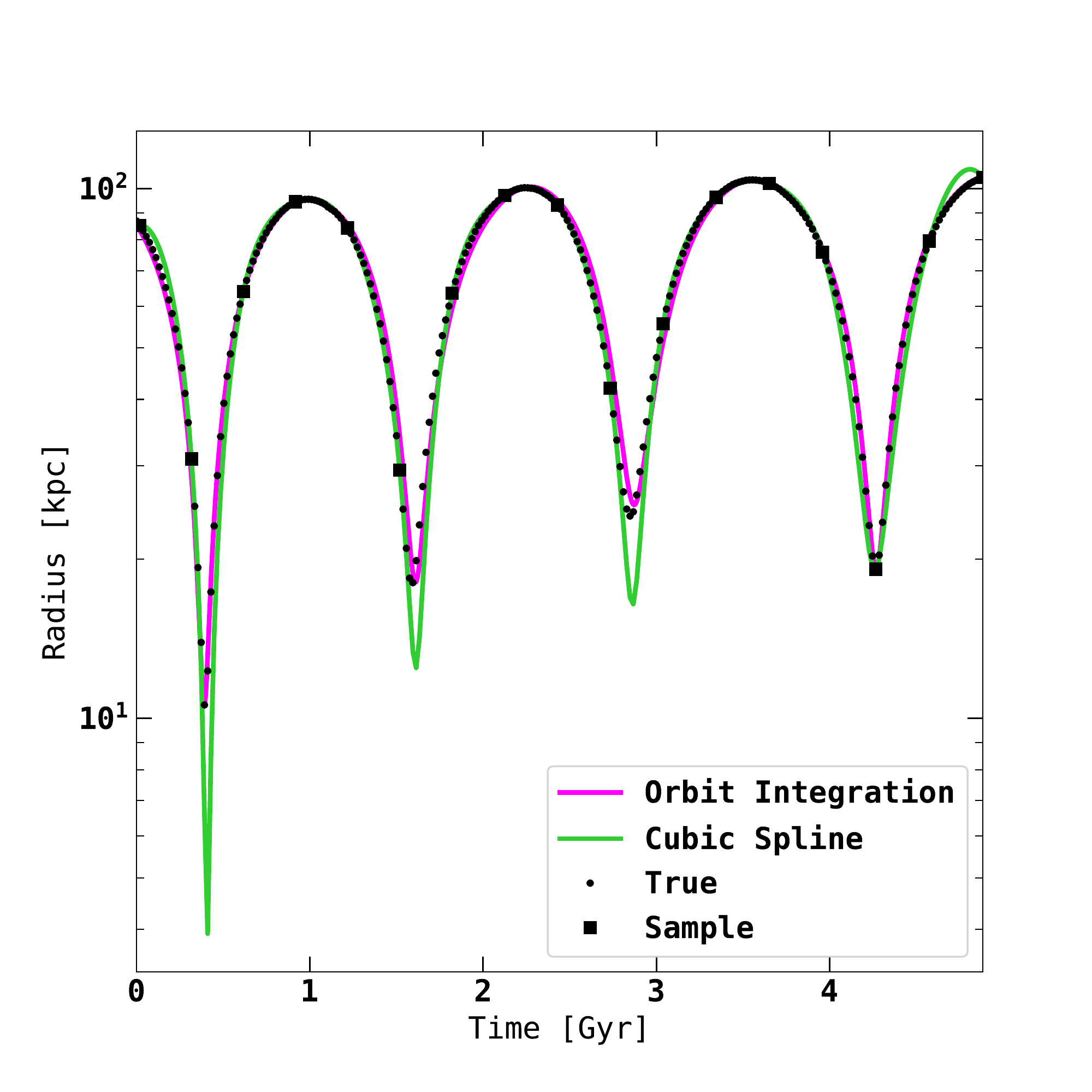}
  \caption{The distance of a subhalo from the halo centre of potential
    over several orbital periods in the Aquarius Aq-A4 simulation. Black
    circles show the distance measured at each snapshot. Black squares
    show the distance at snapshots used for orbit integration and
    fitting cubic splines. The pink line shows the orbit calculated
    using the orbit integration method described above. The green line
    shows the orbit inferred from the cubic spline method introduced
    by \citet{sawala2017}.}
  \label{fig:rad_interp}
\end{figure}

\begin{figure}
  \includegraphics[width=\linewidth]{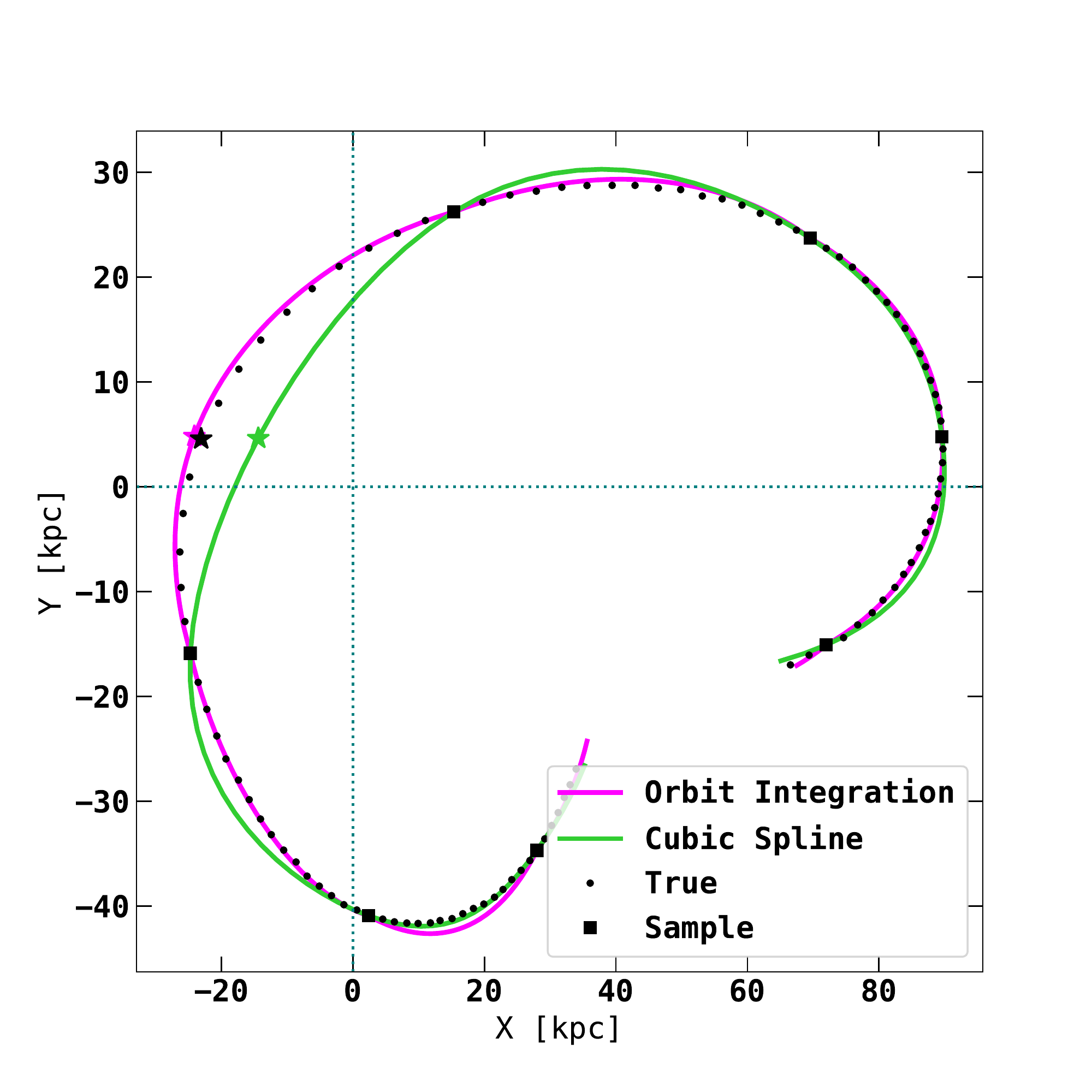}
  \caption{A two-dimensional projection of a portion of the subhalo
    orbit shown in Fig.~\ref{fig:rad_interp}. The orbit is close to
    planar in the $z$ coordinate. The portion of the orbit is chosen
    from the middle of the whole orbit shown in
    Fig.~\ref{fig:rad_interp} to prevent edge effects in the cubic
    spline interpolation. Black circles show the position of the
    subhalo at each snapshot, and black squares the position of
    the subhalo at the snapshots used to calculate the pink (orbit
    integration) and green (cubic spline) lines. Stars show the
    position of the pericentre of the orbit. The black star is almost
    exactly on top of the pink star.}
  \label{fig:x-y_interp}
\end{figure}

To demonstrate the inaccuracies introduced by cubic spline
interpolation, we use the \aquarius simulations. The \aquarius project
is a set of DMO zoom-in simulations of $10^{12}$ \msun dark matter
halos \citep{springel2008}. Specifically, we use the Aq-A4 simulation,
which has 258 snapshots between $z=0.5$ and the present day, 
and a high-resolution particle mass of $3.9\times10^5$
\msun. This time resolution is approximately sixteen times better than
in the \apostle simulations. We select a subset of snapshots with the
same temporal spacing as the snapshots in the \apostle simulations. We can
compare the orbits calculated using the cubic spline interpolation on
the subset of snapshots to the orbit measured in the additional
snapshots not used to fit the cubic splines. Fig.~\ref{fig:rad_interp}
demonstrates how the cubic spline interpolation underestimates the
orbital radius of a subhalo near pericentre. This underestimation
occurs at pericentre as this is where the acceleration experienced by
the subhalo is varying most rapidly. The cubic spline, which assumes
that the acceleration of the subhalo is linear in time between
snapshots, is unable to account for the rapidly varying force acting
on the subhalo as its distance from the centre of the halo changes
rapidly. In Fig.~\ref{fig:x-y_interp} we show a
two-dimensional projection of the orbit over seven snapshots. The
positions plotted for the cubic spline are calculated at the exact
time of the \aquarius snapshots, so any deviations are due solely to
the choice of interpolation.

In contrast, integrating the orbits of subhalos consistently
reproduces the radius and time of pericentre passage with a high
degree of accuracy, as seen in Figs.~\ref{fig:rad_interp}
and~\ref{fig:x-y_interp} .The orbit of the subhalo is determined by
the shape of the potential and its current position and
velocity. Integrating the orbits ensures that the relevant physics is
included, leading to accurate predictions for the positions and
velocities of subhalos between snapshots. On the other hand, the cubic
spline interpolation method has no physical basis, leading to orbits
which do not conserve angular momentum.

\begin{figure}
  \includegraphics[width=\linewidth]{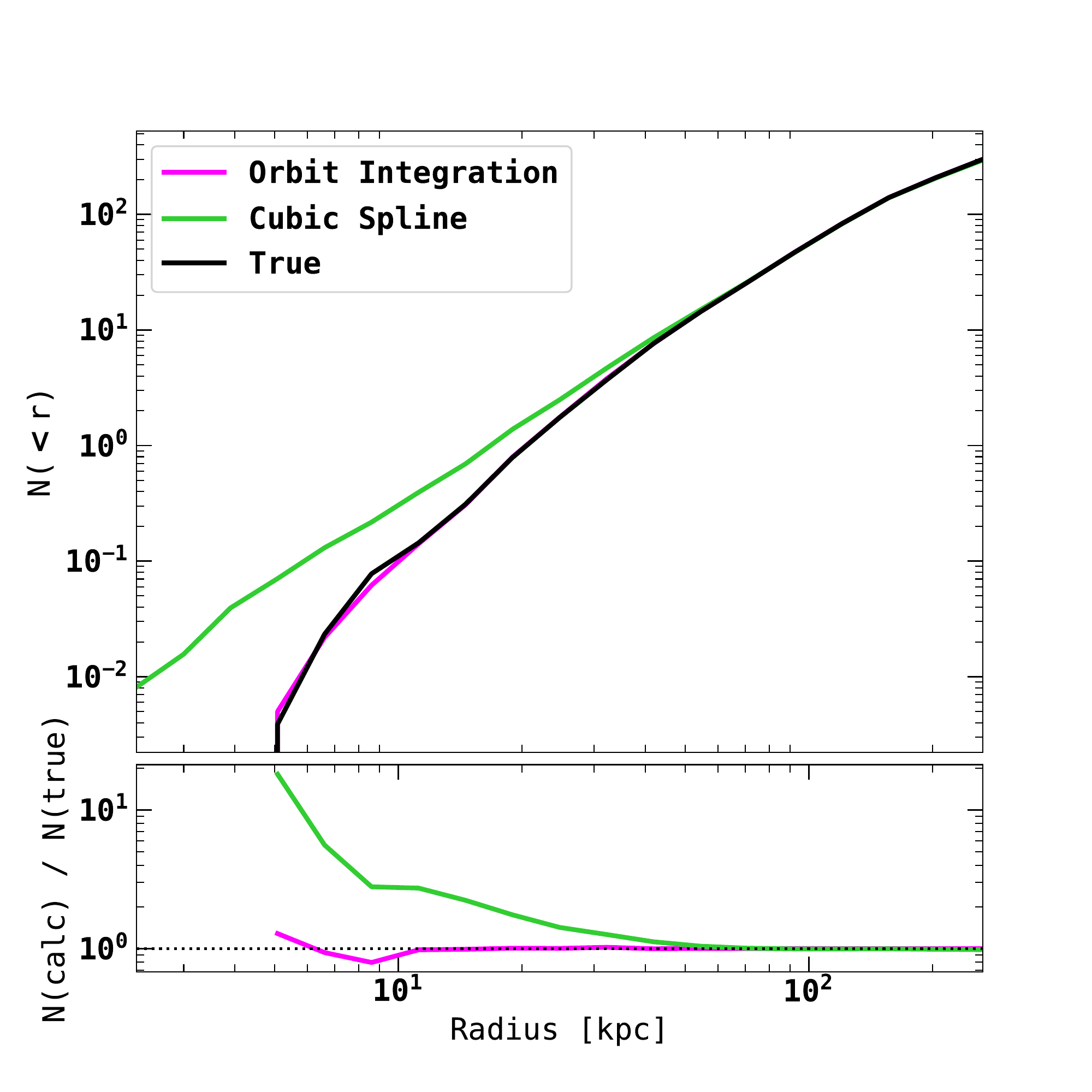}
  \caption{Cumulative radial distribution of subhalos, averaged over a
    5 Gyr period, calculated using orbit integration (pink) and cubic
    spline interpolation (green) on a subset of simulation
    snapshots. The black line shows the radial distribution of
    subhalos measured using all snapshots. The bottom panel shows the
    ratio of the calculated and true number of subhalos inside a given
    radius.}
  \label{fig:numden_interp}
\end{figure}

To quantify the error introduced by the cubic spline interpolation on
the calculation of the subhalo radial distribution, we calculate the
time-averaged cumulative radial distribution of subhalos over a 5 Gyr
period using the positions in the \aquarius snapshots and the
positions calculated using either the cubic spline interpolation or
our orbit integration method.  Fig.~\ref{fig:numden_interp} shows that
the tendency of the cubic spline to underestimate the orbital radius
of a subhalo leads to a large overprediction of the abundance of
subhalos at radii less than 20 kpc (around 10\% of \rvir).  At
distances of less than 5 kpc from the halo centre, the cubic spline
interpolation method predicts a significant chance of observing
substructure despite no object having ever passed so close to the halo
centre. By contrast, the orbit integration method matches the actual
radial distributions perfectly. 

\section{Abundance of substructure in hydrodynamical simulations}

\begin{figure*}
  \includegraphics[width=\linewidth]{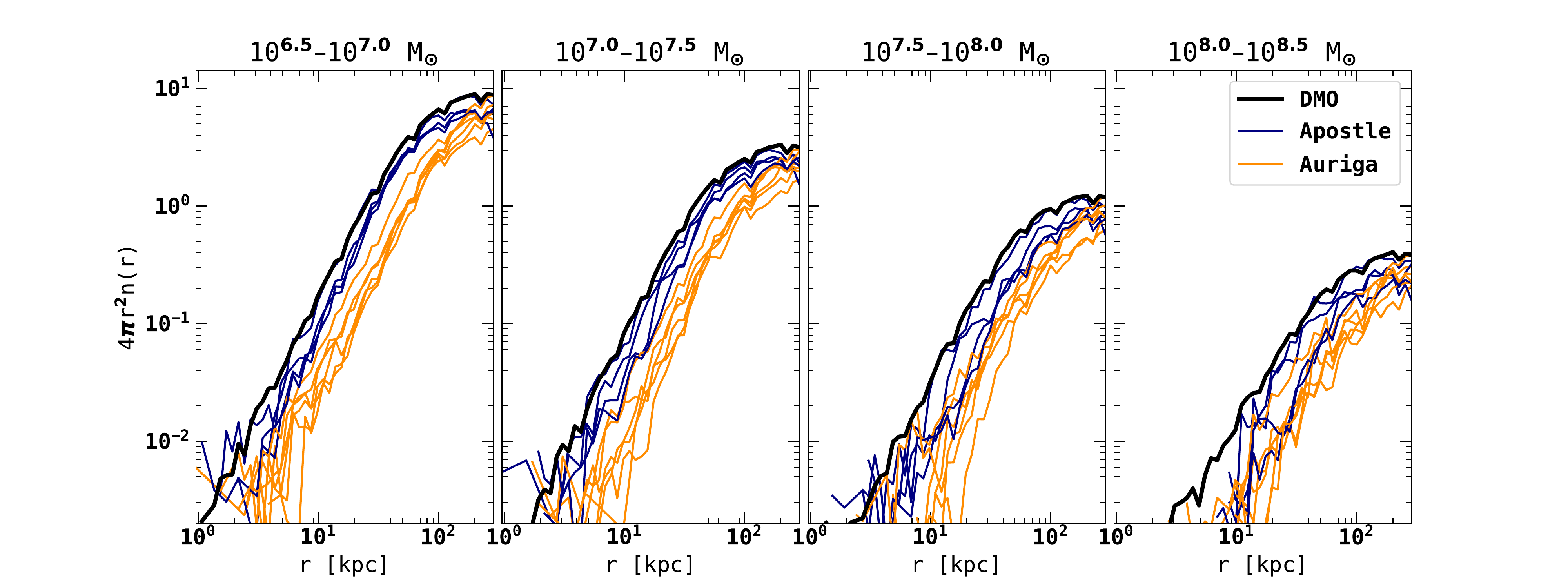}
  \caption{Linear density of subhalos in hydrodynamical and DMO
    versions of the \apostle (blue) and \auriga (orange) simulations as a
    function of radius. The black line shows the median radial density
    of subhalos in all DMO simulations. Each panel corresponds to a
    different subhalo mass bin as indicated. The results are time averaged 
    over a period of 5 Gyr.} 
  \label{fig:abundance}
\end{figure*}

The central galaxies that form in the \auriga simulations are
significantly more massive than those that form in the \apostle
simulations, even though they both have broadly
similar halo masses. We show in this section that the mass of the
galaxy has a marked effect on subhalo abundance, even at distances
well beyond the edge of the galaxy. In Fig.~\ref{fig:abundance} we
compare the radial distribution of subhalos in the \apostle, 
\auriga and DMO simulations. The effect of the larger \auriga galaxies is to
reduce the abundance of subhalos at all radii. We find that the size
of the reduction depends strongly on radius but is broadly independent
of mass for subhalos in the range $10^{6.5}-10^{8.5}$ \msun, in
agreement with the conclusions of \citet{sawala2017}. 

\begin{figure}
  \includegraphics[width=\linewidth]{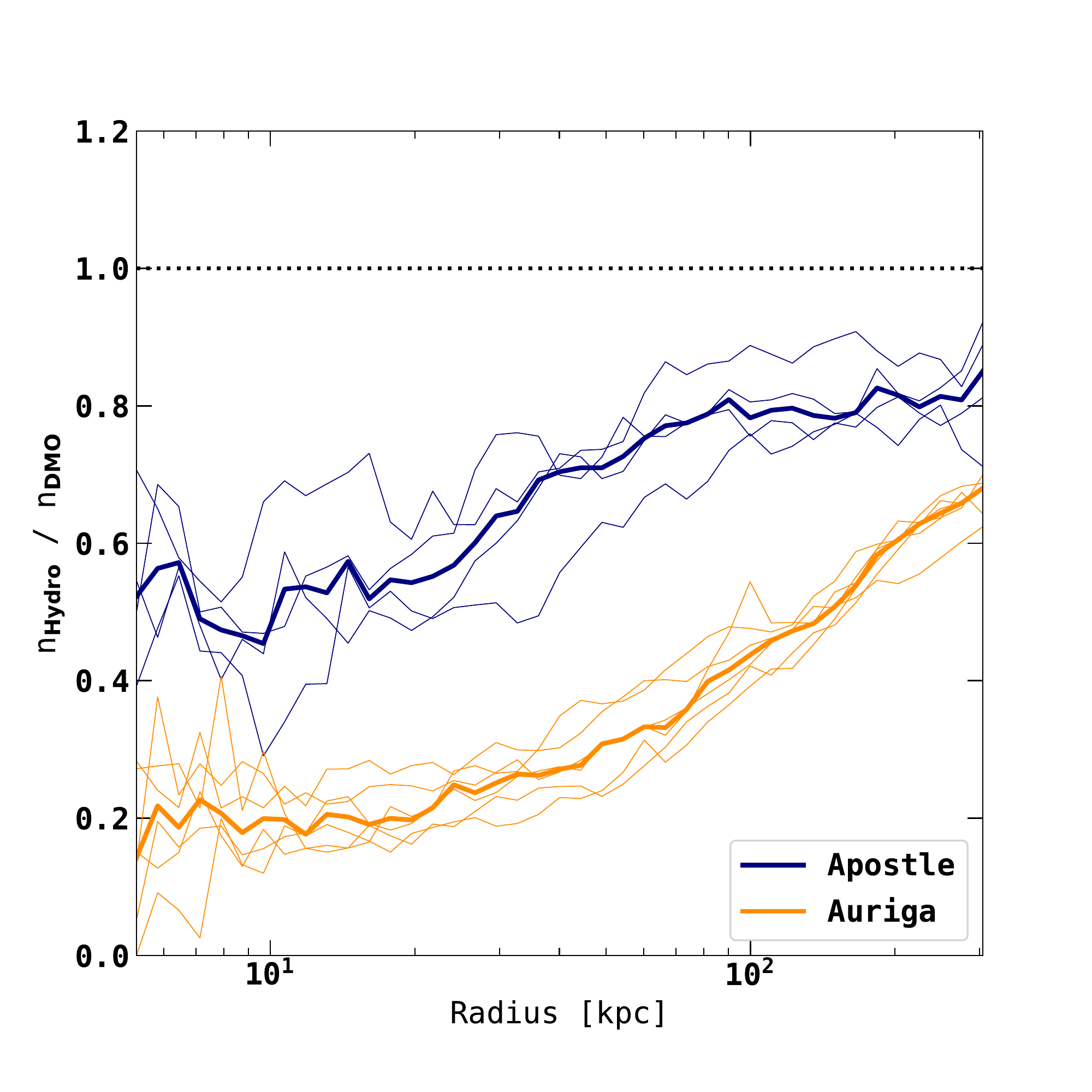}
  \caption{Ratio of the radial number density of subhalos in
    hydrodynamical and DMO versions of the \apostle (blue) and \auriga
    (orange) simulations, for subhalos with masses in the range
    $10^{6.5}-10^{8.5}$ \msun. Thin lines show the reduction in
    subhalo abundance for individual halos and the thick lines the
    median of the thin lines.}
  \label{fig:ratios}
\end{figure}

The reduction in subhalo abundance as a function of radius is shown
explicitly in Fig.~\ref{fig:ratios}. Fundamental tests of the CDM
model, for example using stellar streams to search for substructure, 
are sensitive to substructure within 20~kpc of the centre of
the halo (or equivalently $\sim 10$\% of \rvir for a Milky Way-sized
halo). At these radii, the presence of the galaxy reduces the
substructure abundance by 50\% in the \apostle and by 80\% in the
\auriga simulations relative to the DMO case. The \apostle simulations
predict over twice as many dark (i.e. low-mass) substructures as the
\auriga simulations.

\begin{figure*}
  \includegraphics[width=\linewidth]{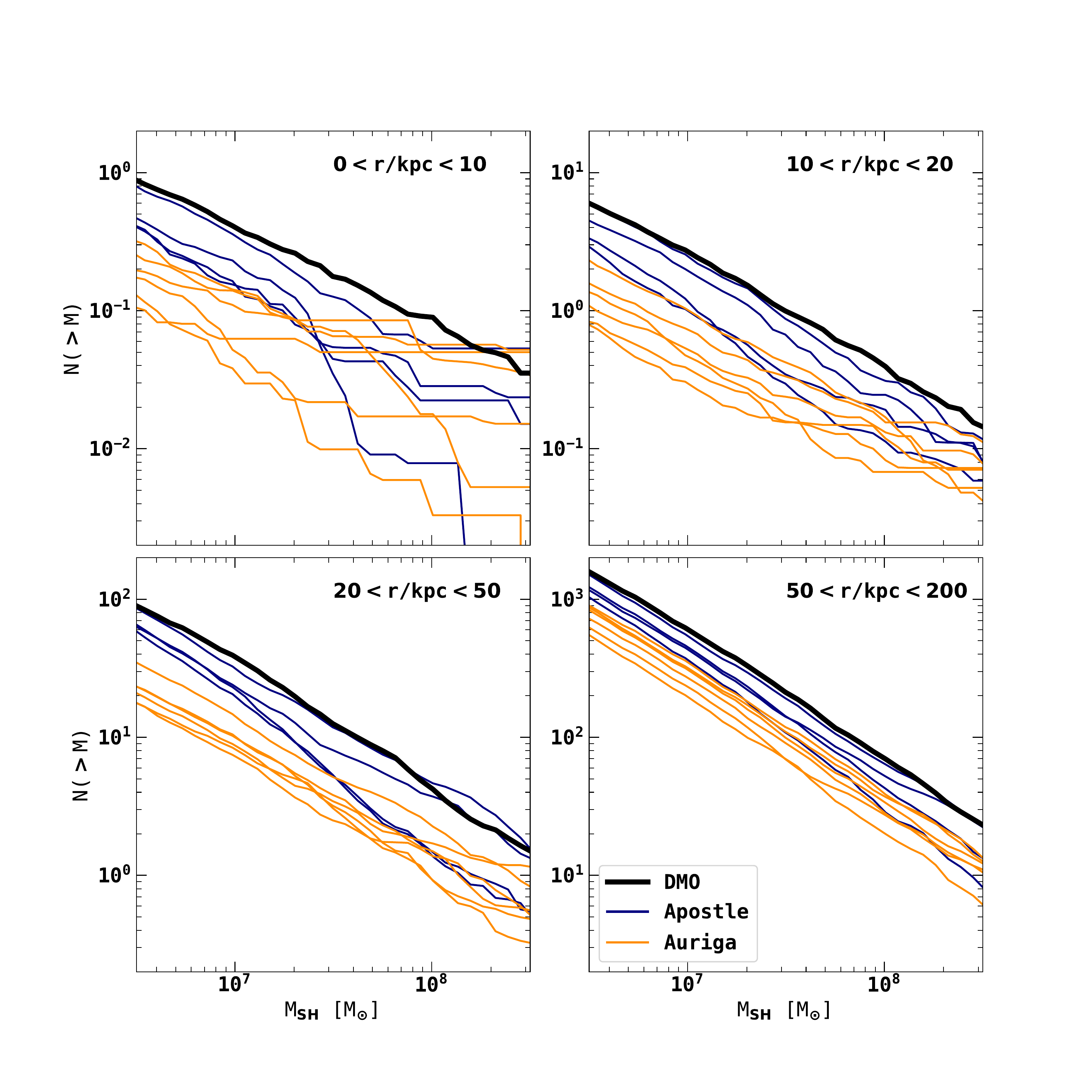}
  \caption{Cumulative subhalo mass functions for subhalos in the
    \apostle (blue) and \auriga (orange) hydrodynamical
    simulations. Each panel represents a different spherical
    shell. The thick black lines show the median cumulative subhalo
    mass function of all \apostle and \auriga DMO subhalos in each
    radial bin.} 
  \label{fig:mass_fn}
\end{figure*}

In Fig.~\ref{fig:mass_fn} we show the cumulative subhalo mass
functions in four spherical shells in the DMO and hydrodynamical 
versions of the \apostle and \auriga simulations. Power-law fits to
the differential mass functions have slopes between -1.8 and -1.9 in
the two outermost shells, consistent with the findings of both
\citet{springel2008} and \citet{sawala2017}. At distances less than
20~kpc from the halo centre (top panels) we find that the slopes of
the mass functions in the \auriga hydrodynamical simulations are
significantly shallower than the corresponding slopes in
\textsc{Apostle}, suggesting that the implementation of baryonic
physics in \auriga leads to a more pronounced reduction of small-mass
relative to high-mass halos. This is simply because less massive halos are more prone to tidal disruption, rather than any systematic difference between the orbital distributions of smaller and larger halos.  We fit the median mass function in each
radial bin with a power law using a non-linear least squares
method. In the innermost radial bin the slope in the \auriga
hydrodynamical simulations is -1.4, whereas the slope in \apostle simulations
is -1.7. Values for all power-law fits are listed in
Table~\ref{tab:mf_fits}.

\begin{table}
\begin{tabular}{l|cccc}
              & 0-10 & 10-20 & 20-50 & 50-200 \\ \hline
Apostle DMO   & -1.74    & -1.88     & -1.90     & -1.91      \\
Auriga DMO    & -1.69    & -1.77     & -1.92     & -1.93      \\
Apostle Hydro & -1.73    & -1.86     & -1.92     & -1.93      \\
Auriga Hydro  & -1.44    & -1.64     & -1.82     & -1.94     
\end{tabular}
\caption{Power-law slopes for differential subhalo mass functions in the mass 
  range $(10^{6.5}-10^{8.5})$~\msun in DMO and hydrodynamical
  simulations, in four spherical shells. The width of the spherical
  shell (top row) is given in kpc.} 
\label{tab:mf_fits}
\end{table}

\subsection{Subhalo abundance far from the central galaxy}
\label{LargeRadii}

As the distance from the central galaxy increases, the reduction in subhalo abundance
caused by the inclusion of baryonic physics should asymptote
to a constant value, at a radius where the tidal field of the central galaxy has
no impact on the evolution of small halos.
This is indeed what we see in Fig.~\ref{fig:ratios} which
shows that the ratio of subhalo abundance in the hydrodynamical and DMO
simulations rises with distance from the centre of the halo, until it
begins to plateau at a radius of $\sim 300$ kpc. The reason that the 
lines in Fig. \ref{fig:ratios} do not plateau at a value of 1 is because
we computed the number density of subhalos in fixed mass bins, without 
correcting for the effects described in \S\ref{MassCorrection}.
The radius at which the reduction in subhalo
abundance plateaus to a constant value is significantly outside of 
\rvir, which has a typical value of 220 kpc for the halos in our sample. 
Thus, the impact of the central galaxy seems to extend surprisingly far, well
beyond the extent of the galactic disks.
This conclusion is based on two observations.

\begin{figure}
  \includegraphics[width=\linewidth]{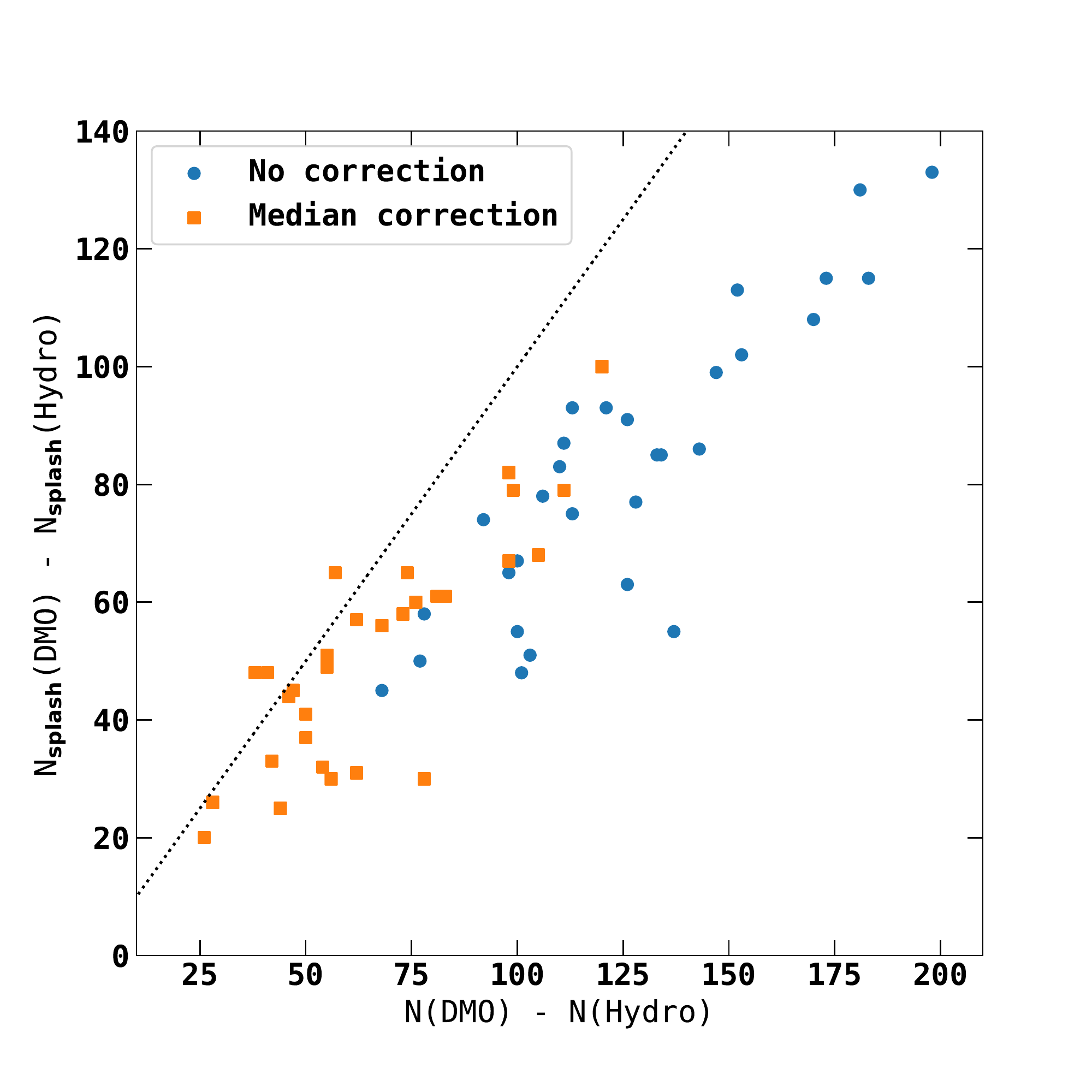}
  \caption{Difference in the number of subhalos with masses in the
    range $(10^7-10^8)$~\msun in a spherical shell of width
    200--300 kpc in the DMO and hydrodynamical simulations plotted
    against the difference in the number of subhalos in this shell
    that have previously been inside 70\% of \rvir (labelled by the
    subscript $splash$). Each point represents a halo from the level~4
    \auriga suite of simulations. The blue points show the halos where
    we have not applied any correction to the masses of DMO
    subhalos. The orange points show the result when we correct the
    masses of DMO subhalos using the mass correction technique
    described in \S\ref{MassCorrection}.}
  \label{fig:splash_test}
\end{figure}

Firstly, the velocity anisotropy of subhalos is lower (implying more
circularly-biased orbits) in the hydrodynamical simulations compared
to the DMO case. The difference in velocity anisotropy only becomes
negligible at around 300 kpc. Secondly, the number of halos that have
been in and out of the main halo is larger in the DMO than in the
hydrodynamical simulations.  To show this, we count the number of
halos of mass $(10^7-10^8)$~\msun in a spherical shell between 200
and 300 kpc. For each subhalo we check if it has previously
been close to the central galaxy\footnote{We adopt a radius of $0.7\times$\rvir as our 
definition of `close'. This is a typical 
radius at which the tidal forces from the spherically averaged mass 
distributions in the hydrodynamical simulations are equal to the 
tidal forces in the DMO simulations.}. We find that the difference
in the number of subhalos between the hydrodynamical and DMO
simulations is strongly correlated with the difference in the number
of subhalos that have been close to the centre of the halo. In other
words, at large radii there exists a population of DMO subhalos that
have fallen into the halo, survived their passage through the centre,
and reemerged. These are sometimes called ``splashback halos'';
\citealt{gill2005}.  Many of their hydrodynamical counterparts do not
survive the encounter with the galaxy at the centre of the halo, and
so we observe the abundance ratio continuing to rise to distances of 300 kpc from 
the centre of the halo, well beyond \rvir. The results of this calculation 
for the level~4 suite of \auriga simulations (see \S\ref{Methods}) are shown in
Fig.~\ref{fig:splash_test}. Here we compare results with and without
the mass correction described in \S\ref{MassCorrection}. We see a
clear correlation in both cases; however, when the mass correction is
applied, the points fall roughly along the expected 1:1 line.

To explore this point further, we compare the evolution of a
population of subhalos matched between the DMO and hydrodynamical
simulations. We match the subhalos firstly by particle IDs, using the
matching criteria of \citet{bose2017} and, secondly, by requiring that
the subhalos should have the same mass and distance from the centre of
the main halo to within 10\%. This criteria is quite restrictive, and effectively
limits our sample to objects which have not yet had an interaction with the
main halo yet. Matched objects have the same orbits at redshift $z=1$, so
we can confidently attribute present day differences between the matched objects
to interactions that occur during the time period we study.
We track the masses and positions of our matched sample between redshift $z=1$
and the present day, a period of roughly 8 Gyr and compare matched subhalos that 
were between 200 and 300 kpc from the centre of the halo at redshift $z=1$. From this
sample we select objects that ceased to exist before redshift $z=0$ in
the hydrodynamical version but that survive to the present day in the
DMO version. Subhalos that meet this criteria are approximately three
times as common as subhalos that survive in the hydrodynamical
simulation but are destroyed in the DMO simulation.

\begin{figure}
  \includegraphics[width=\linewidth]{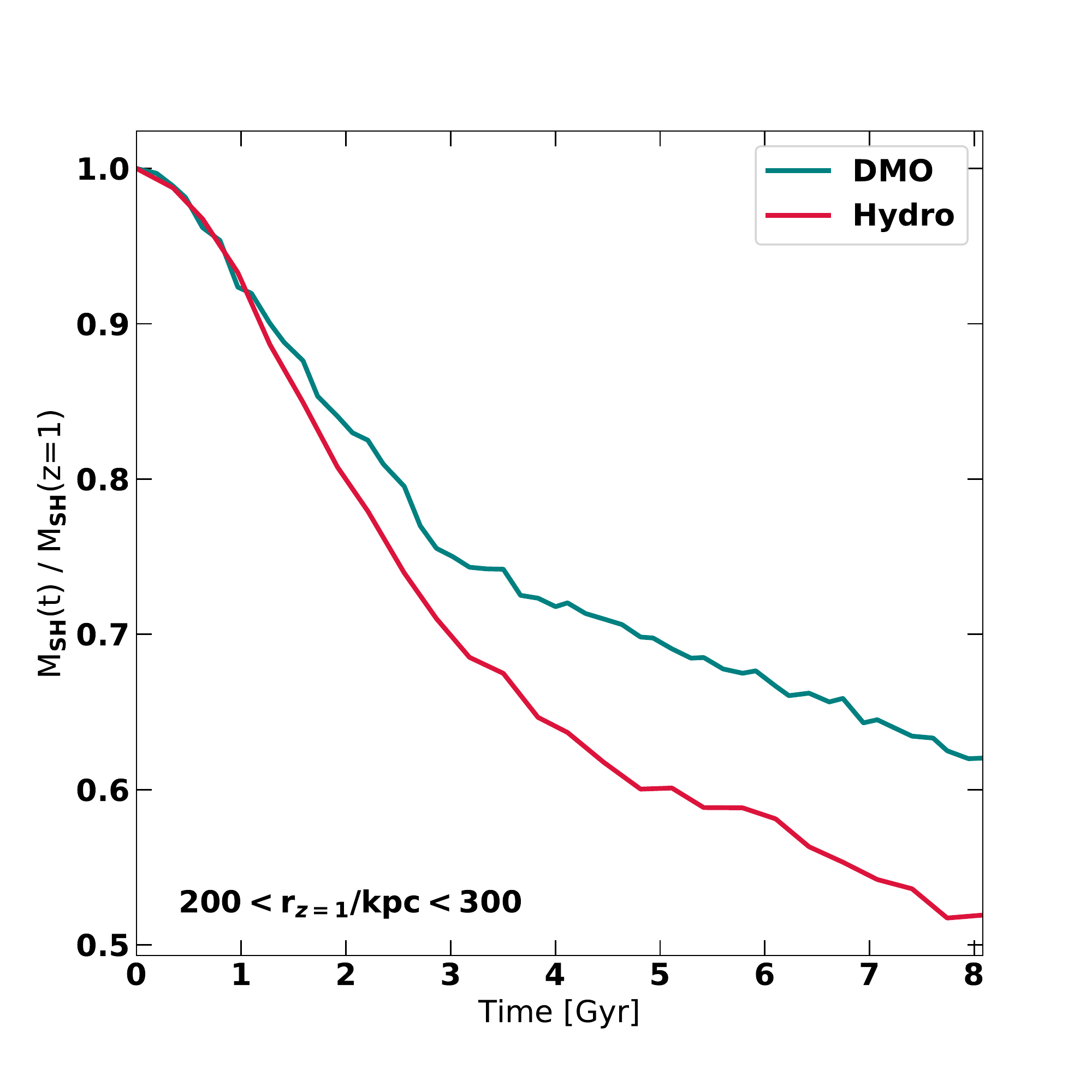}
  \caption{Mass loss of subhalos between redshift, $z=1$, and the
    present day for a sample of subhalos matched between DMO and
    hydrodynamical versions of the \auriga simulations. Subhalos are
    selected  at $z=1$ to be between 200-300~kpc from the centre of
    the main halo and to have a mass in the range $(10^{6.5}-10^{8.5})$~\msun. Each
    line shows the median reduction in mass as a function of time for
    subhalos which survive to the present day.} 
  \label{fig:massloss}
\end{figure}

We can also use our matched sample of subhalos to assess the role of
mass stripping (rather than complete destruction) in the reduction in
subhalo abundance. The steepness of the subhalo mass function means
that it is possible to measure a reduction in subhalo abundance in a
particular mass bin, if subhalos undergo significant stripping without
any destruction taking place at all. We select a sample of matched
subhalos that lie between 200 and 300~kpc from the centre of the halo
at $z=1$, have a mass in the DMO simulation in the range
$(10^{6.5}-10^{8.5})$~\msun and survive to the present day. Although
we do not specify it in advance, we find that the dynamics of these
matched objects, specifically their average distance from the centre
of the main halo as a function of time, is identical in the
hydrodynamical and DMO samples. Fig.~\ref{fig:massloss} shows the
median reduction in subhalo mass as a function of time. DMO subhalos
lose an average of 38\% of their mass, whilst subhalos in \auriga lose
an average of 49\%. Thus, a halo in the hydrodynamical simulation with
the same initial mass as its DMO counterpart at redshift $z=1$ will
be, on average, about 20\% less massive today, even if it shares the
same radial distance history. This merely reflects the enhanced tidal
stripping in the later case due to the presence of the massive central
galaxy. We can quantify the contribution of this effect to the
overall reduction in abundance as follows.

We assume a power-law mass function of the form,
\begin{equation}
    \frac{dN}{dM_0}=kM_0^{\alpha} \;,
\end{equation}
where $M_0$ is the uncorrected mass. The mass of the subhalo after
stripping is given by $M_1=\beta M_0$. Thus,
\begin{equation}
    \frac{dN}{dM_1}=k\beta^{-\alpha-1}M_1^{\alpha} \;,
\end{equation}
so the ratio of the mass functions is given by
$\beta^{-\alpha-1}$. Taking values of $\alpha=-1.9$ for the power law
slope of the subhalo mass function\citep{springel2008},
and $\beta=0.8$ for the stripping factor (the difference in stripping
between the hydrodynamical and DMO simulations) gives a value
of 0.82 for the ratio of the mass functions, corresponding to an 18\%
reduction in the number of objects. This stripping effect is the
dominant cause for the reduction in subhalo abundance for distances
greater than 200 kpc from the centre of the halo. Stronger stripping
in the hydrodynamical simulations also explains why the orange points
(i.e. with corrected masses) in Fig.~\ref{fig:splash_test} lie slightly
below the 1:1 line on average.

\begin{figure}
  \includegraphics[width=\linewidth]{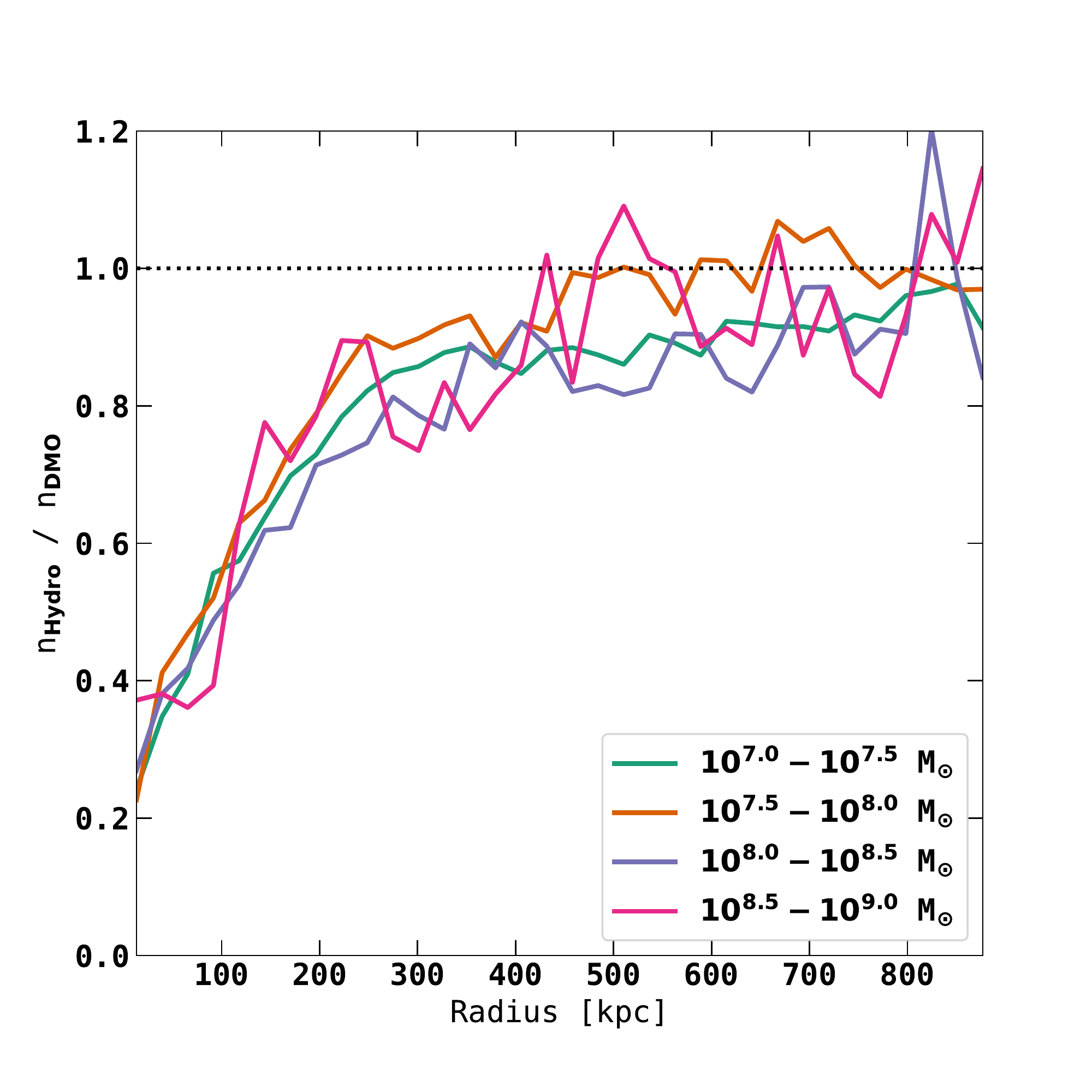}
  \caption{The ratio of the radial number density of subhalos in
    hydrodynamical to DMO versions of the \auriga simulations. Each
    line represents a different mass bin as indicated in the
    legend. The results are time-averaged over a period of 5 Gyr. The
    masses of subhalos used in this figure are the ``effective''
    masses, calculated using the method described in \S
    \ref{MassCorrection}.}
  \label{fig:ratio_far}
\end{figure}

Finally, in Fig.~\ref{fig:ratio_far} we show the reduction in subhalo
abundance in the hydrodynamical relative to the DMO versions of the
\auriga simulations out to a radius of 800 kpc. Each line gives the
median reduction in subhalo abundance in a given mass bin for the six
high-resolution \auriga halos. In this plot we multiplied the masses
of DMO subhalos by 0.75 following the method described in
\S\ref{MassCorrection}. We see a clear change in gradient around
300~kpc from the main halo. Inside this distance the reduction in
abundance decreases approximately linearly with radius. Even at
distances of 400~kpc or greater from the halo, the ratio of number
densities has not yet reached unity. This is partly a result of the
inability of our mass correction to capture the full range of
different growth histories of subhalos in the DMO and hydrodynamical
simulations, and partly a result of the same processes that destroy
subhalos near the centre of the main halo being played out in smaller
halos that later merge into the main one. For the \auriga suite of
simulations there are, on average, 3 galaxies with stellar mass greater
than $10^8$ \msun between 400 and 800 kpc from the centre of the main
halo, as well as dozens of smaller galaxies. The presence of massive
galaxies (and their more disruptive tidal forces) at the centre of all
these halos/subhalos will also contribute in a small way to the
reduction in the abundance of substructure relative to the DMO version
of the simulations.

\section{Subhalo velocities}

An accurate estimate of the expected velocity distribution of low mass
substructures is a critical input into methods to search for
small-mass dark substructures from measured gaps in cold stellar
streams. In this section we examine the velocity distributions in our
simulations; contrasting the two sets, we can gain some insight into
the size of the  theoretical uncertainties in these distributions. This
topic has been explored already by, for example, \cite{sawala2017}.

\begin{figure*}
  \includegraphics[width=\linewidth]{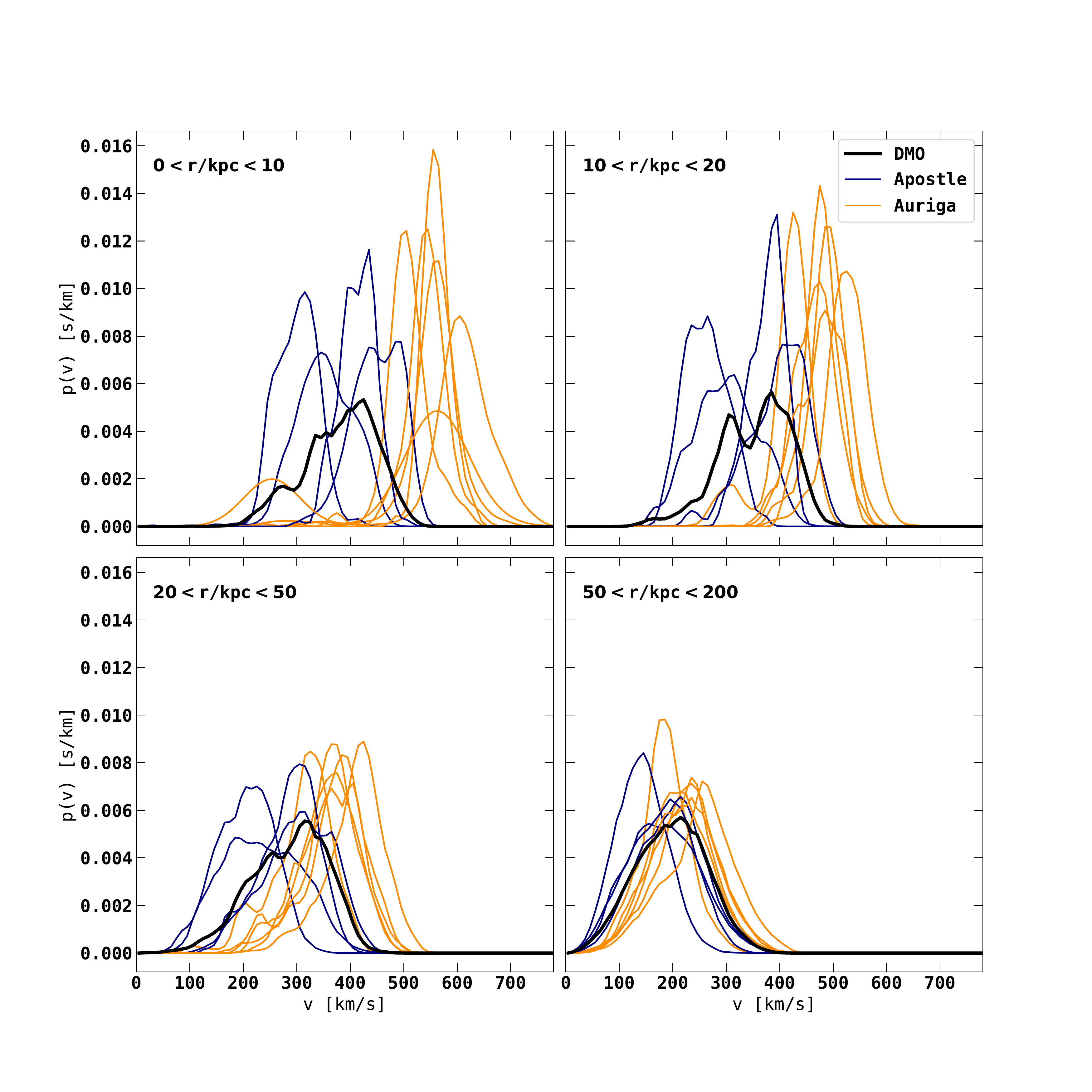}
  \caption{Probability distributions of the speed (relative to the
    host halo) in spherical shells for subhalos of mass in the range
    $(10^{6.5}-10^{8.5})$~\msun in the \apostle (blue) and \auriga
    (orange) simulations. Thick black lines show the median velocity
    distribution of all DMO subhalos in each bin.}
  \label{fig:modv}
\end{figure*}

To obtain a robust estimate of the velocity distributions, we employ
kernel-density estimation \citep{rosenblatt1956, parzen1962}, using a
Gaussian kernel and applying Scott's rule to estimate the bandwidth
\citep{scott2015}. The distribution of subhalo speeds as a function of
radius is shown in Fig.~\ref{fig:modv}. The presence of the central
galaxy affects the distributions relative to the DMO case for both
hydrodynamical sets of simulations.  The impact of the more massive
central galaxies in the \auriga simulations is clear. The depth of the
potential well is larger, leading to a greater radial acceleration as
subhalos fall inwards. This effect of the central galaxy is also
manifest in the \apostle simulations but it is much weaker reflecting
the smaller masses of the central galaxies.  We note that no such
effect was observed by \citet{sawala2017}, probably as result of
inaccuracies in their interpolation scheme.

We find that the distribution of subhalo speeds is generally well fit
by a Rician distribution, in agreement with \citet{sawala2017}. The
Rician distribution is a two-parameter model given by 
\begin{equation}
    f(x\mid\nu,\sigma) = \frac{x}{\sigma^2}\exp\left(\frac{-(x^2+\nu^2)}
{2\sigma^2}\right)I_0\left(\frac{x\nu}{\sigma^2}\right),
\end{equation}
where $I_0$ is the modified Bessel function of the first kind with
order zero. The $\nu$ parameter controls the location of the peak,
with a value of 0 giving a Maxwellian distribution. The $\sigma$
parameter controls the width of the distribution.  The parameters of
the fits are given in Table~\ref{tab:rician_fits}.
\begin{table}
\begin{tabular}{l|cccc}
              & 0-10 & 10-20 & 20-50 & 50-200 \\ \hline
Apostle DMO   & 351, 110    & 310, 101     & 242, 97     & 167, 78      \\
Auriga DMO    & 384, 68    & 355, 66     & 285, 75     & 191, 71      \\
Apostle Hydro & 379, 75    & 326, 83     & 249, 81     & 165, 71      \\
Auriga Hydro  & 554, 48    & 480, 43     & 356, 56     & 211, 63     
\end{tabular}
\caption{Values of the parameters $\nu$ and $\sigma$ obtained from
  fitting a Rician distribution to the median values of the
  velocity distributions shown in Fig.~\ref{fig:modv}, in km/s. Each
  column correspond to a different radial bin, with the width of the
  shell in kiloparsecs.}  
\label{tab:rician_fits} 
\end{table}

\begin{figure*}
  \includegraphics[width=\linewidth]{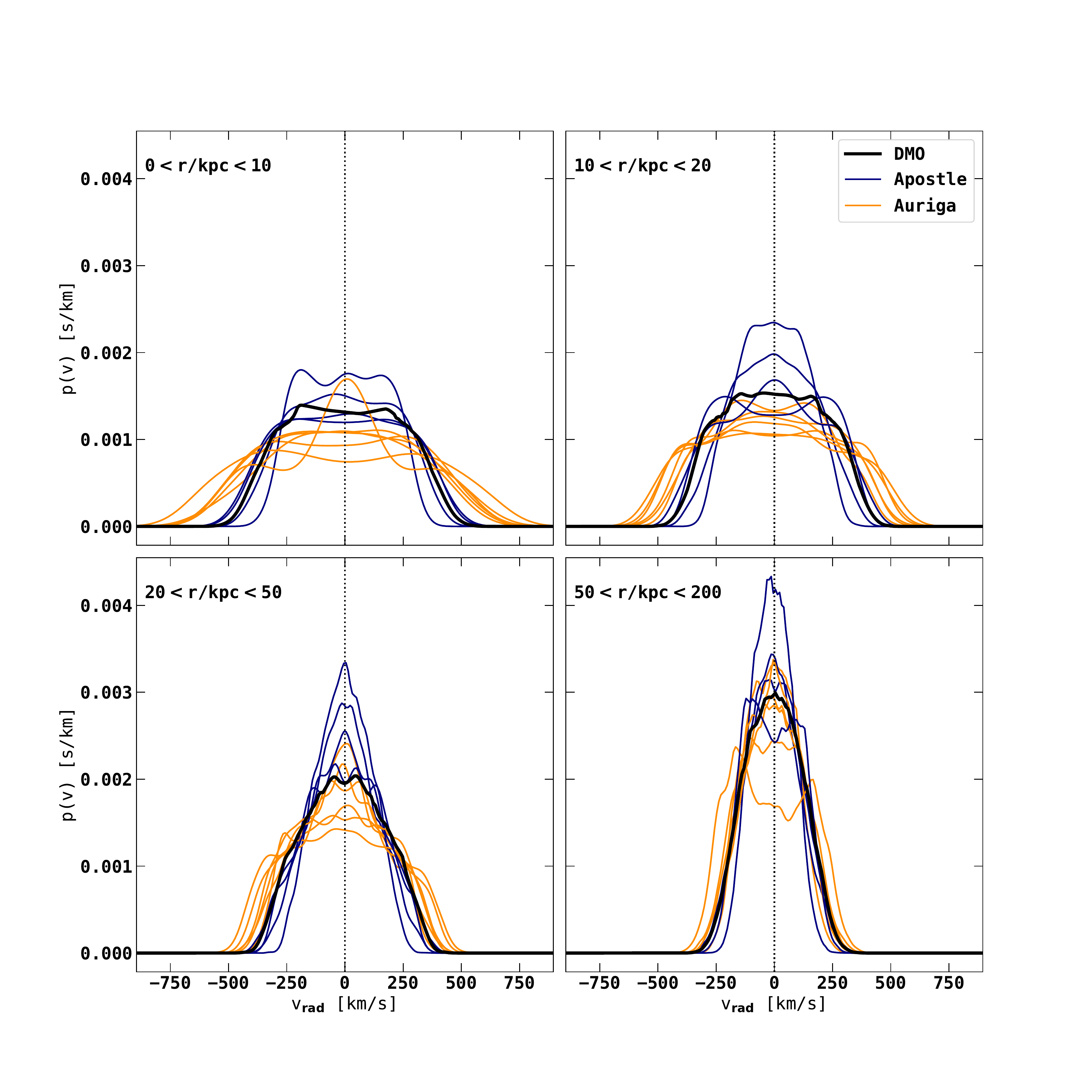}
  \caption{Probability distributions of radial velocities (relative to
    the host host halo) in spherical shells for subhalos with masses
    in the range $(10^{6.5}-10^{8.5})$~\msun in the \apostle (blue)
    and \auriga (orange) simulations. Thick black lines show the
    median velocity distribution of all DMO subhalos in each bin.}
  \label{fig:vrad}
\end{figure*}

The distributions of subhalo {\em radial} velocities in the same
radial bins used in Fig.~\ref{fig:modv} are shown in
Fig.~\ref{fig:vrad}. \citet{sawala2017} found that close to the halo
centre, the distribution of subhalo radial velocities in the Apostle
simulations was well described by a double
Gaussian. Fig.~\ref{fig:x-y_interp} shows how plunging orbits
calculated using the cubic spline interpolation method pass closer to
the centre of the halo than the true orbits, with a velocity that is
predominantly tangential during most of the passage through the
central region. This is a general feature of orbits constructed using
the cubic spline interpolation method. Consequently, the dearth of
low-radial velocity orbits reported by \citet{sawala2017} is an artifact of
their orbit reconstruction method. This explains why the velocity
distributions that we find in the top left panel of
Fig.~\ref{fig:vrad} do not show such a pronounced dip around
$v_{\mathrm{rad}}=0$. In the 50--200 kpc radial bins, we see that
one of the \auriga systems has an unusually bimodal velocity distribution.
This distribution is the result of an interaction with another halo between
redshift $z=0.5$ and the present day. A population of subhalos belonging to
the passing halo have flown in and out of the edge of the halo, resulting in
a peak of the negative radial velocity whilst infalling, and a peak in the
positive radial velocity distribution after pericentre.

We can see in Fig.~\ref{fig:vrad} that the deeper gravitational
potential the in hydrodynamical simulations relative to the DMO case
leads to a broadening of the radial velocity distribution, with the
effect being most pronounced in the \auriga simulations at small
radii. This effect is a combination of a greater radial acceleration
and the preferential disruption of objects on more circular orbits
near the centre of the halo. We also note that the distributions are
remarkably symmetrical, even in the outermost spherical shell. This
shows that the subhalo abundance at all radii reflects a balance
between inflow and outflow.

\section{Conclusions}

The large number of small-mass halos predicted by N-body simulations
to form in a $\Lambda$CDM universe provide a key test of the
paradigm. In practice, however, the clear-cut predictions from N-body
simulations are only part of the answer, as some of these small
halos that fall into larger ones can be destroyed by tidal forces
whose strength depends on the contents of the halo, particularly the
galaxy at the centre. Thus, rigorous predictions for the abundance of
subhalos requires modelling the baryonic processes that lead to the
formation of the galaxy. In this paper, we have investigated how the
abundance and velocity distribution of small-mass subhalos
($\sim 10^{6.5}-10^{8.5}$\msun) within galaxy-size halos is affected
by baryon processes and we have compared two different implementations
of such processes using the independent \apostle and \auriga
simulations. 

Since subhalos are quite rare near the centre of the host halo and are
poorly sampled in the limited number of available simulation outputs,
to study their orbits we have integrated the orbits of subhalos between
snapshots, using the publicly available code \textsc{Galpy}.
The results we present are obtained by averaging over a
lookback period of 5~Gyr.

We find that the abundance of substructures is significantly affected
by the way in which baryon processes are treated. At 10\% of the
virial radius, \rvir, the abundance of low-mass substructures is
reduced relative the dark matter-only (DMO) simulations by around 50\%
and 80\% in the \apostle and \auriga simulations respectively. We also
find differences in the slope of the subhalo mass function and the
width and peak location of the velocity distributions, all of which
can be explained by the different masses of the galaxies that form at
the centre of the halos in the two simulations. The more massive
central galaxies in \auriga result in larger tidal forces, which cause
enhanced destruction and stripping of substructures.  Perhaps
surprisingly, we find that the abundance of subhalos in the
hydrodynamical simulations is still lower than in the DMO simulations
even well beyond the virial radius of the halo, particularly in {\sc
  auriga}. This happens because objects that spend the majority of
their orbit far from the central galaxy have highly radial orbits
which take them past the virial radius; some of the objects that
emerge unscathed from the DMO simulation, are destroyed in the
hydrodynamical counterpart.

A deeper potential also causes subhalos to accelerate more as they
move towards the centre of the halo, leading to an increase in the width
of the radial velocity distributions. We also find that the peak of
the distribution of subhalo speeds is shifted to significantly higher
values in the hydrodynamical simulations, with the largest changes
occurring near the centre of the \auriga simulations.

\citet{sawala2017} and \citet{garrison2017} investigated similar
processes to those we have studied here, the former using the same
\apostle simulations that we too have analyzed. Our results differ
significantly from theirs. We have shown that this is because the
cubic spline method they used to interpolate orbits between snapshots
is insufficiently accurate to follow the orbits near the centre of the
halo.  Our orbit integration method predicts less substructure at
small distances from the halo centre. We also do not observe the
velocity biases described by \citet{sawala2017}. However, we find that
the conclusion of \citet{sawala2017} that objects on radial orbits are
more likely to undergo disruption by the central galaxy holds true.

Roughly speaking the \apostle and \auriga simulations bracket the
range of theoretical uncertainty for the abundance and velocity
distribution of substructures near the centre of a galaxy like the
Milky Way. \apostle underpredicts the mass of the Milky Way by factors
of 2-3, whereas, on average, the \auriga galaxies overpredict it by
factors of 1.5-2.  The halo-to-halo variations in the velocity
distributions is smaller than the differences seen in our two
hydrodynamical simulations. This size of theoretical uncertainty is
eminently reducible by improved modelling of the baryonic physics of
galaxy formation.

\section{Acknowledgements}
CSF acknowledges support from European Research Council (ERC) Advanced
Investigator grant DMIDAS (GA 786910). This work was also supported by
the Consolidated Grant for Astronomy at Durham (ST/L00075X/1). It made
use the DiRAC Data Centric system at Durham University, operated by
the Institute for Computational Cosmology on behalf of the STFC DiRAC
HPC Facility (\url{www.dirac.ac.uk}). This equipment was funded by BIS
National E-infrastructure capital grant ST/K00042X/1, STFC capital
grants ST/H008519/1 and ST/K00087X/1, STFC DiRAC Operations grant
ST/K003267/1 and Durham University. DiRAC is part of the National
E-Infrastructure.

\bibliographystyle{mnras}
\bibliography{destruction}

\end{document}